# Using Jupyter for reproducible scientific workflows

**Marijan Beg**
Faculty of Engineering and Physical Sciences, University of Southampton, University Road, SO17 1BJ Southampton, United Kingdom

**Juliette Taka**
Logilab, 104 Boulevard Auguste Blanqui, 75013 Paris, France

**Thomas Kluyver**
European XFEL, Holzkoppel 4, 22869 Schenefeld, Germany

**Alexander Konovalov**
School of Computer Science, University of St Andrews, Jack Cole Building, North Haugh, KY16 9SX St Andrews, United Kingdom

**Min Ragan-Kelley**
Simula Research Laboratory, Martin Linges vei 25, 1364 Fornebu, Norway

**Nicolas M. Thiéry**
Laboratoire de Recherche en Informatique, Université Paris-Saclay, CNRS, 91405 Orsay, France

**Hans Fangohr**
Max Planck Institute for the Structure and Dynamics of Matter, Luruper Chaussee 149, 22761 Hamburg, Germany
Center for Free-Electron Laser Science, Luruper Chaussee 149, 22761 Hamburg, Germany
European XFEL, Holzkoppel 4, 22869 Schenefeld, Germany
Faculty of Engineering and Physical Sciences, University of Southampton, University Road, SO17 1BJ Southampton, United Kingdom

*Abstract*—Literate computing has emerged as an important tool for computational studies and open science, with growing folklore of best practices. In this work, we report two case studies – one in computational magnetism and another in computational mathematics – where domain-specific software was exposed to the Jupyter environment. This enables high level control of simulations and computation, interactive exploration of computational results, batch processing on HPC resources, and reproducible workflow documentation in Jupyter notebooks. In the first study, Ubermag drives existing computational micromagnetics software through a domain-specific language embedded in Python. In the second study, a dedicated Jupyter kernel interfaces with the GAP system for computational discrete algebra and its dedicated programming language. In light of these case studies, we discuss the benefits of this approach, including progress towards more reproducible and re-usable research results and outputs, notably through the use of infrastructure such as JupyterHub and Binder.

■ **INTRODUCTION**

Research usually results in a publication that presents and shares the obtained findings and conclusions. For a publication to be scientifically valid, it must present the methodology rigorously, so that readers can follow the "recipe" and reproduce the results. If this criterion is met, the publication is considered reproducible. *Reproducible* publications are more easily *re-usable* and thus provide a significant opportunity to make (often tax-payer funded) research more impactful. However, the reproducibility of computational work is usually hindered not only by a lack of data or meta-data but also by a lack of details on the procedure and tools used:

1) The source code of the software used is not available.
2) Information on the computing environment, such as the hardware, operating system, supporting libraries, and (if required) code compilation details is not revealed.
3) The exact procedure which led to the results reported in the publication is not shared. This should include the set of parameters used, the simulation and data analysis procedure, and any additional data cleaning, processing, and visualization. Ideally, these are shared as open-source code and analysis scripts used to perform the simulation and to read, analyze, and visualize the resulting data. This way, the entire process can be repeated by re-running simulation and/or analysis scripts. A human-readable document detailing the computational steps taken, despite being "better-than-nothing", is still insufficient to ensure reproducibility, and keeping a detailed log of all steps taken during a computational study is often impossible.

Reproducibility is a challenging question and spans a range of different topics. In this work, we focus on one of them. We describe the features and capabilities of the Jupyter environment that, in our view, make it a highly productive environment for computational science and mathematics, while facilitating reproducibility.

The topic of *bitwise* reproducibility is outside the scope of this work: even with the same hardware and same software, it may be difficult to reproduce computational results to be bitwise identical. This can originate from the non-associativity of floating-point operations combined with parallel execution or from compiler optimizations. Bitwise reproducibility is not always required to be achieved.

In the last decade, literate computing has emerged as an important tool for computational studies and open science, with an ever-growing set of best practices. In this paper, we review and expand some of these best practices in the context of two case studies: computational magnetism and mathematics. This is based on the experience of enabling and applying Jupyter environments in these fields as a part of the OpenDreamKit (https://opendreamkit.org/) project.

To be able to run computational studies from the Jupyter environment, it is necessary to either have the simulation and/or analysis code exposed to a general-purpose programming language supported by Jupyter, or have a dedicated Jupyter kernel for the computational libraries. Although the main topic of this work is the overview of features and capabilities of the Jupyter environment for reproducible workflows, we begin by discussing how a computational library can be exposed to Jupyter as a necessary prerequisite.

## Prerequisite: Exposing computational libraries to the Jupyter environment

Computational studies often use existing computational (legacy) tools. These could be executables called from the command line or libraries that are used within a programming language. For the approach suggested here, these computational tools need to be accessible to scientists from a general-purpose programming language supported by Jupyter (such as Python). For some domains, such as pure mathematics research, there are domain-specific languages with enough power to be used directly as the programming language in notebooks (e.g., Singular and GAP). In other areas, exposing computational tools to a general-purpose programming language is the key to integrating them into researchers' custom code. A key benefit of making computational tools available in a general-purpose programming language is that the computation can be driven flexibly using the control structures provided by that language. For example, a simulation can conveniently be repeated with a range of parameters through a for-loop, rather than having to change a configuration file for each value and trigger execution of the simulation manually.

Making the computational capability accessible from a general-purpose programming language supported through a Jupyter kernel such as Python may be trivial – for example, if the required code is already a Python library. When the computational functionality is locked into an executable, one can create an interface layer so that functionality can be accessed via a Python function or class [1]: input parameters will then be translated into configuration files, the executable called, outputs retrieved, and finally, the results returned.

If the computational tool uses a programming language that Jupyter does not support, another possibility is to implement a Jupyter kernel for that language so that the computational library can be exposed to the Jupyter environment (as done for GAP and SageMath for example).

Over time, scientific communities tend to accumulate functions and classes that are used repeatedly, and occasionally, through organic changes or a systematic restructuring of those computational capabilities, a *domain-specific language* is created, which is embedded in a general-purpose programming language such as Python. Depending on the design of this language, its existence and joint use by researchers of that domain can help to unify and improve computational tasks in the community, avoid duplication of work, support transfer of knowledge and reproducibility. Examples of such domain-specific languages include Ubermag in magnetism, SageMath in pure mathematics, and the atomic simulation environment in chemistry [2].

## Features of the Jupyter research environment

Project Jupyter is a set of open-source software projects for interactive and exploratory computing emerging from IPython. The central component offered by Jupyter is the Jupyter Notebook – a web-based interactive computing platform. It allows users to create data- and code-driven narratives that combine live (re-executable) code, equations, narrative text, interactive dashboards, and other rich media. Jupyter Notebook documents provide a complete and executable record of a computation that can be shared with others in a way that has not been possible before [3]. Within the Jupyter Notebook, all libraries available in Python can be imported and combined flexibly. Other languages (such as Julia, R, Haskell, Bash, and many more) are supported through other Jupyter Notebook kernels. In this work, we suggest using a Jupyter research environment from which computational studies can be driven and conducted efficiently. In this section, we discuss the benefits of using the Jupyter environment for reproducible scientific workflows.

1. One study – one document

The notebook allows us to carry out an entire study within a single notebook and provides a complete and executable record of the process. It is possible to put the interpretation of the results into the same document, immediately below the graphical, tabular or text-based output that needs to be described. The "one study – one document" approach has immediate advantages:

- Scientists can be more efficient as they do not have to search for parts of the study (scripts, data files, plots) when trying to understand the data and authoring the associated paper.
- The study is more easily reproducible (see item 6, below).

However, putting all the code, data, and narrative into a single notebook could substantially affect the notebook's readability. Thus, it is necessary to decide which parts of the code should be in libraries and imported in the notebook.

2. Easily shareable

Jupyter notebooks can be converted to other file formats, such as HTML, LaTeX and PDF. This is useful because someone working on a notebook can share it with collaborators, supervisors, or management without asking them to install any additional software.

3. Interactive execution or as batch job

Using a Jupyter notebook often involves interactively editing it, executing cells, inspecting computed outputs, modifying commands, and re-executing, while understanding the computational research question. Once a useful processing sequence has been found, the researcher often wants to repeat that, potentially with different input data. For such scenarios, a notebook can be executed from the command line (using the

`nbconvert` tool), treating the notebook like a script or a batch job. As the notebook executes in batch mode, it computes the output cells, including images and other multimedia, as if it were executed interactively, and the outputs are stored into the notebook file for later analysis and inspection. Execution of notebooks as a script is a convenient way to use the computational power of a high-performance computing facility where such notebook jobs can be submitted to the batch queue.

Where input data needs to be varied, two solutions are available: `nbparameterise` and `papermill`. With these tools, assignments in the first cells of a notebook can be modified before the notebook is executed as a script.

### 4. Static and interactive software documention

Writing research software documentation is a particular challenge in academia. Small teams may not see the need to document their research code, as they can learn about it directly from one another.

Jupyter notebooks offer an efficient method for creating documentation. The popular *Sphinx* documentation software can use Jupyter notebooks as the documentation source with the `nbsphinx` plugin, and create HTML and PDF documents. Demos and tutorials written in notebooks can complement reference documentation in Sphinx's default *reStructuredText* input format. Notebooks have several benefits for extended examples in documentation:

- It takes less time to create documentation as the author can type commands and explanations into the same document, and the outputs that the commands produce (text and images) appear immediately in the notebook.
- After changing the user interface or computational algorithms, re-executing the documentation notebooks will often show where the documentation needs changing.
- Tools like `nbval` can automatically re-execute the notebooks and raise test errors if the execution fails, or the computed outputs have changed. This means continuous integration can be used to check the documentation and warn developers if changes in the code affect the illustrated behaviour.

- Using Binder, the documentation notebook can be executed interactively by the user (see item 5, below).

### 5. Executable interactive documents in the cloud (Binder)

The open-source Binder project [4] and Binder instances such as myBinder offer customized computational environments in the cloud on-demand, in which notebooks can be executed interactively. To use the free myBinder service, one needs to create a publicly readable git-repository containing Jupyter notebooks and a specification of the software required to execute these notebooks. This specification follows existing standards, such as a Python-style `requirements.txt` file, conda `environment.yml` file, or `Dockerfile`. The myBinder service is invoked when a URL is requested containing the path to the GitHub repository. The myBinder service searches that repository for the software specification, creates a suitable container, adds a Jupyter server to the container, and exposes that server to the user. Figure 1 offers an artistic illustration of a typical scenario for using Binder in the research workflow. Other use cases include:

- Providing a computational environment for workshops or teaching purposes: participants are given the URL to invoke the service, and are presented a Jupyter session, in which they find the notebooks the presenter/teacher has prepared. No software installation (other than having a modern web browser) is required for participants.
- Providing interactive documentation: Given Binder-compatible specifications, documentation can be presented as an executable notebook through myBinder, allowing the person reading the documentation to interactively explore the software's behavior, for example by repeatedly modifying and executing commands provided as part of the documentation.
- Demonstrating and disseminating small computational studies: Jupyter notebooks can be used to document computational processes. For example, for dissemination or to demonstrate reproducibility, as we explain in item 6, below.

The related *Voilà* project can execute note-books (for example on myBinder) and hide all code cells, making an interactive dashboard to display and explore data without the source code.

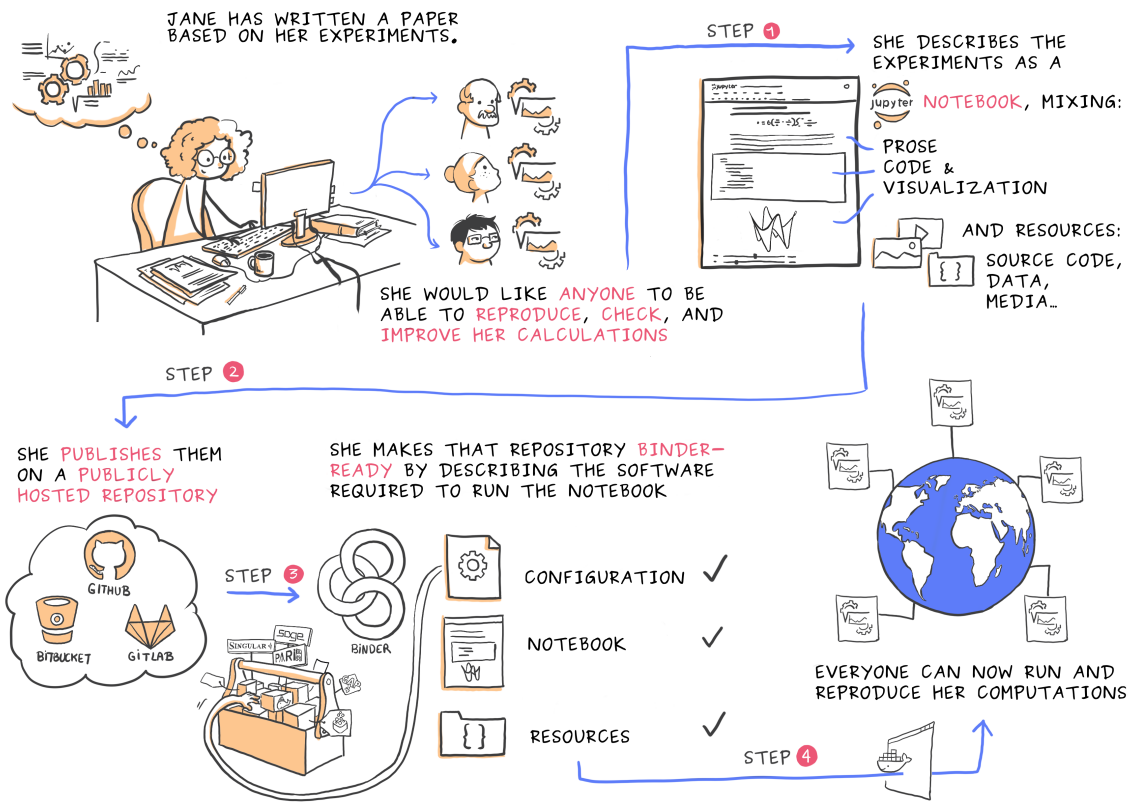

**Figure 1.** An artistic depiction of a scenario in which a researcher shares her computational workflow with others in the Jupyter environment, taking advantage of the Binder project. Licensed under "Creative Commons Attribution Share Alike 4.0 International" – Juliette Taka and Nicolas M. Thiéry. Publishing reproducible logbooks explainer comic strip. Zenodo. DOI: 10.5281/zenodo.4421040 (2018).

6. Reproducibility – combining data, code, and software environment

Reproducibility of scientific results is a cornerstone of our interpretation of science: only results that can be reproduced are accepted as proven insight. We see an emerging trend that journals and research councils increasingly (and justifiably) ask for details on how published results can be reproduced, or at least expect authors to provide that information if a reader requests.

It is often impossible to truly document an entire computational workflow, software requirements, hardware used, and other parameters within a conventional manuscript submission. The Jupyter-based research environment can help because it makes the process of publishing reproducible computational results easily achievable:

- The "one document – one study" model automatically records all parameters, processing commands, and outputs, demonstrating the process leading to the result obtained with that notebook. By sharing the notebooks in a public repository, a DOI can be assigned via Zenodo to preserve the repository's content permanently and make it citable.
- Notebooks that create central figures and statements of publications will likely need underlying libraries. To re-execute the notebook, we need a way to specify a computational environment containing these libraries and Binder provides that possibility. Although specifying exact versions of underlying libraries is recommended, Binder does not guarantee that this would lead to the same computational environment at any point in the future, and therefore, it cannot entirely address the issue of so-called software collapse where the underlying libraries and interfaces become deprecated, compilers and compiler optimization methods change, etc.
- By publishing the notebooks reproducing central results together with software environment specifications for Binder in an open repository, anyone with Internet access and a browser can inspect and re-execute these notebooks and thus reproduce the publication.

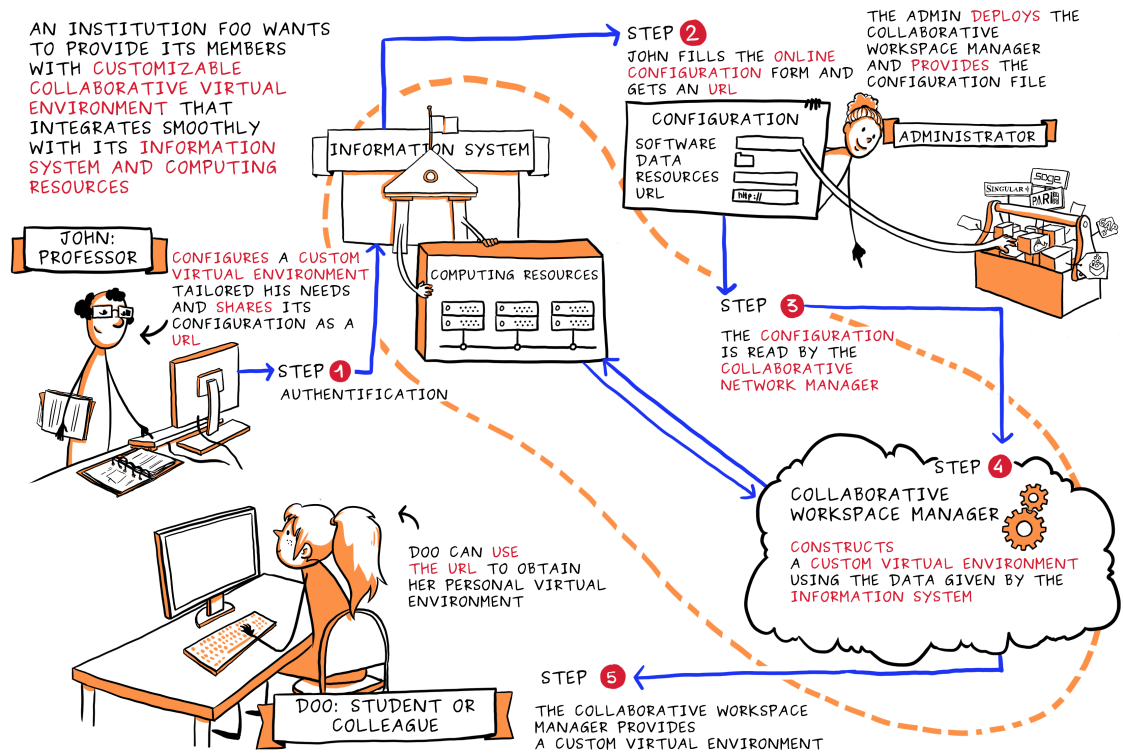

**Figure 2.** An artistic illustration of a configurable JupyterHub where a lecturer provides a customized software environment to support their teaching. JupyterHub can be accessed and used through a web browser and does not require local installation of any software. Institutional computing and storage resources are used, and users have to authenticate themselves. Licensed under "Creative Commons Attribution Share Alike 4.0 International" – Juliette Taka and Nicolas M. Thiéry. On demand customizable Virtual Environments with JupyterHub explainer comic strip. Zenodo. DOI: 10.5281/zenodo.4432267 (2019).

A key benefit of being able to reproduce a publication in this way is that the study can be modified and extended easily: *reproducibility enables re-usability*. This can provide efficiency gains for science overall as it allows scientists to focus on new insights rather than having to spend time re-creating known knowledge as a starting point of their new study.

## 7. Remote access to institutional compute resources – JupyterHub

The discussion above assumed that notebooks were running on the user's computer. The JupyterHub software allows institutional provision of Jupyter Notebook services. It allows users of an institution to authenticate with their organizational credentials and access a Jupyter environment running on the institution's infrastructure. Typically, any files and folders the user is allowed to access will also be made available to them through JupyterHub, including access to shared data and folders where they can save their notebooks.

The institution generally predefines the soft-ware environment in which the notebook server executes. However, the technology is available to use the software specification as for Binder to create a customized computing environment on-demand. A vital point of the user experience is that only a web browser is required to access the JupyterHub and to carry out computational work using these resources remotely. Figure 2 shows an artistic illustration of the scenario where an instructor works with their institution to provide students with a customized software environment. Other use cases of JupyterHub installations include research facilities and universities providing access to their (high-performance) computing resources through Jupyter notebooks, where traditionally `ssh` or remote desktops may have been used.

## 8. Blending script and GUI-driven exploration methods

The `IPyWidgets` Jupyter extension provides selection menus, sliders, radio buttons, and other GUI-like graphical interaction widgets to Jupyter notebooks. The Notebook allows embedding such graphical widgets inside the notebook, and users can combine the usual scripted analysis with activation of such widgets where desired. They can be used, for example, to vary the input parameter values and explore a data set or computational results. Although less reproducible than typed commands, widgets can be useful for rapid feedback on different possibilities.

## 9. Potential disadvantages

Above we focused on the features and capabilities of the Jupyter research environment to support computational workflows in science. Here, we want to discuss some downsides that have come up either in our work or as feedback from users of Jupyter-based computational tools we developed.

*(a) Undefined notebook state*
Top-to-bottom arrangement of cells in a notebook implies that they should be executed in that order. One of the Jupyter Notebook's key features is that the code cells can be executed in an arbitrary order – the user can select (and modify) any cell and then execute it. This can be useful while exploring a data set or a property of computation, or even to debug the cell's code. The execution order used in a notebook is not stored when the notebook is saved. Therefore, it is critical to remember that, by executing cells out of order, we may create different results from when we execute all cells in order.

There is a practical solution to this. When the exploratory phase is completed, the best practice is to restart the kernel to ensure the notebook's state is forgotten and then execute all cells from top to bottom. This ensures that the results in the notebook are obtained from running the cells in order, and this version of the notebook should be saved and shared.

*(b) Opening Jupyter Notebook*
Among the feedback we receive from some users who come across Jupyter notebooks for the first time is that the way a Jupyter server is started is "strange". Users who are not used to the command prompt may find it unusual to open an application that way, instead of "double-clicking".

*(c) Rapid development of Jupyter ecosystem*
Improvements to Project Jupyter and the surrounding software ecosystem appear at a rapid rate. For instance, for the issues described in items (a) and (b), contributions providing solutions have already emerged, and there is no space here to introduce more of the multitude of high-productivity tools that have been created. It is challenging to follow all the developments and find the most appropriate tool for a given task. Conferences such as JupyterCon help disseminate new contributions and help to avoid duplication of development efforts.

*(d) Sustainability of myBinder.org*
Since 2016 (and at time of writing), a federation of Binder instances is operated as a service available on the world wide web at `mybinder.org`. The federation is operated by the Jupyter team, in collaboration with the Turing Institute and GESIS (Leibniz Institute for Social Sciences). Computing resources are sponsored by Google Cloud, OVHCloud, the Turing Institute, and GESIS. The federation serves approximately 25,000 Binder instances on a typical weekday, with the Google Cloud instance serving approximately 70% of this traffic. These sponsorships are mostly renewed annually and can result in members of the federation halting the operation due to periods without funding. We hope that the sustainability of the Binder federation will improve if more financially-stable members join, for example, as a part of the European Open Science Cloud initiative.

## Case studies

### Computational magnetism

Computational magnetism complements theoretical and experimental methods to support research in magnetism. For example, it is used to develop sensors as well as data storage and information processing devices. It is used both in academia and industry to explain experimental observations, design experiments, improve device and product-designs virtually, and verify theoretical predictions.

The Object-Oriented MicroMagnetic Framework (OOMMF) [5] is a micromagnetic simulation tool, initially developed during the 1990s at the National Institute of Standards and Technology (NIST). It solves non-linear time-dependent partial differential equations using the finite-difference method. It is probably the most widely used and most trusted simulation tool in the computational magnetism community. It was written in C++, wrapped with Tcl, and driven through configuration files that follow the Tcl syntax.

The typical computational workflow the user must follow to simulate a particular problem is to write a configuration file. After that, the user runs OOMMF by providing the configuration file to the OOMMF executable. When the OOMMF run is complete, results are saved in OOMMF-specific file formats. Finally, the user analyzes the result files.

One of the specific goals of a computational micromagnetic study is parameter-space exploration. More precisely, the user repeats the simulation for different values of input parameters by changing them in the configuration file. It is often difficult to automate this, and it is challenging for the user to keep a log of all steps performed in the entire micromagnetic study. Besides, postprocessing and analysis of results is performed outside OOMMF, using techniques and scripts that are mostly developed by the user, or carried out manually. Consequently, it is hard to track, record, and convey the exact simulation procedure. Without this information, resulting publications are generally not reproducible.

To address this situation, we developed a Python interface to the OOMMF executable. This allows us to conduct computational magnetism simulations from within the Jupyter notebook to capitalize on the benefits of this environment.

We developed a set of Python libraries we refer to as Ubermag, which expose the computational capabilities of OOMMF so that it can be controlled from Python. These Python libraries provide a domain-specific language to define a micromagnetic problem [1]. A micromagnetic model, defined using the domain-specific language, is not aware of the particular simulation tool that will perform the actual micromagnetic simulation, and it is only used to describe the model. When a simulation is required, the model is translated into the OOMMF configuration file, the OOMMF executable is called, and the output files are read. By exposing the micromagnetic simulation capabilities to Python and driving the research from Jupyter Notebook, we have available all the benefits of the Jupyter research environment.

To demonstrate the use of Ubermag, we use standard problem 3 as an example. Standard problem 3 is a standardized problem posed by the micromagnetic community to test, validate, and compare different simulation tools. It describes a magnetic cube of edge length $L$ with two different magnetization states that can occur as local energy minima, called the flower state and the vortex state. The main question of standard problem 3 is "For what edge length $L$ have the flower state and the vortex state the same energy?"

In the conventional OOMMF workflow, it is necessary to run the micromagnetic simulations for different edge lengths and different initial magnetization states. After every simulation, the total energy is recorded and saved within a tab-separated data file. Finally, one extracts the magnetic energy values from all the saved files and plots them as a function of edge length for both magnetization states. From the plot, an estimation of the energy crossing would be made.

By using our Python interface to OOMMF integrated into a Jupyter notebook, we can loop over different input parameters to obtain this crossing in a plot. Furthermore, we can make use of the Python scientific stack, in particular, a root-finding method such as `bisect` from `scipy`. A Jupyter notebook solving standard problem 3 can be found in the repository accompanying this work [M. Beg *et al.* Using Jupyter for reproducible scientific workflows. GitHub: https://github.com/marijanbeg/2021-paper-jupyter-reproducible-workflows, DOI: 10.5281/zenodo.4382225 (2021)]. We show the two most relevant code cells inside the Jupyter notebook in Figure 3.

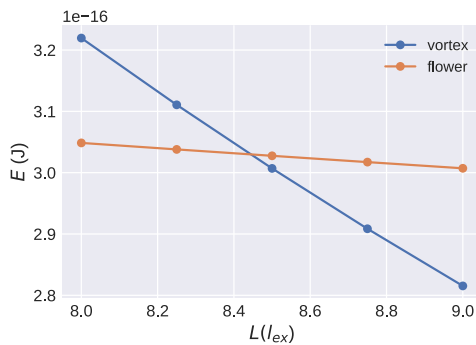

**Figure 3.** Running computational magnetism simulations through Python in a Jupyter notebook allows the use of the Python scientific stack and results in a self-contained record combining narrative, code, and results.

Ubermag and the Jupyter environment simplify the efforts to make computational magnetism publications reproducible. For each figure in the publication, one notebook can be provided (find examples in Refs. [6], [7]). Using Binder, the community can inspect and re-run all the calculations in the cloud and make the publication reproducible.

## Computational studies in mathematical research

Many of the leading open-source mathematical software systems (including GAP, LinBox, PARI/GP, OSCAR, SageMath, and Singular) have been made inter-operable with the Jupyter ecosystem through bespoke or general-purpose kernels (C++, Python, Julia, ...). Focusing on one of these systems, for the sake of concreteness, we illustrate how this supports sharing and publishing reproducible computational studies in mathematical research together with the underlying research code.

GAP is an open-source system for discrete computational algebra, with particular emphasis on computational group theory. It is used routinely by mathematicians in these fields and beyond to support teaching and research, notably through computational exploration. It provides a domain-specific language, also called GAP, and a runtime system with a command-line interface. It can also be used as a library by other systems such as SageMath or OSCAR.

GAP has been developed for decades by a community of researchers, teachers, and research software engineers. It has an established mechanism for user-contributed extensions, called *packages*, which may be submitted for the redistribution with the system, and a formal refereeing process. The current release of GAP (4.11.0) includes 152 packages that serve different purposes, from providing data libraries and extending the system's infrastructure for testing and writing documentation, to adding new functionality and sharing research codes that underpin their authors' publications. The latter scenario may require specific expertise and motivation from a working mathematician who uses GAP, and not everyone will be able to invest efforts in sharing their code in this way. Furthermore, it is not always justifiable to organize a supplementary code for a paper as a new GAP package. Instead, authors can combine Jupyter research environments with additional services and parts of the infrastructure for GAP packages to share reproducible computational studies while following good code development practices from the start.

Let us illustrate this with the publication in Ref. [8], which presents a polynomial-time algorithm for solving a major problem in computational group theory, which remained open since 1999 [9]. An essential addition to the paper is the author's GAP implementation of the algorithm. The authors published this implementation in the publicly hosted repository. At once, this ensures long term archival through the Software Heritage project, and with a small additional step, it makes it citable through Zenodo. The repository contains an interactive narrative document – a Jupyter notebook using the GAP Jupyter kernel [10] – combining text, mathematics, inputs, and outputs, and may even be viewed as a slideshow (one could, of course, have separate notebooks for different purposes).

Following best practices for organizing reproducible computational studies (see e.g., Ref. [11]), the code is not written in the notebook itself but loaded from external source files. These are text files that can be easily managed with version control, reused from multiple Jupyter notebooks, and tested using the GAP automated testing setup. Also, the authors made the repository *Binder-ready*. Any user (e.g., readers or referees of the paper) can run the notebook and reproduce its execution on Binder itself or – with additional expertise to install the required assets – on their own computing resource. To achieve this, the authors followed the template in Ref. [12], which also brings in continuous integration to automatically check the code against several past, current, and development releases of GAP, and produce coverage reports on how thoroughly the tests exercise the code. It boils down to creating a `tst` directory with the test files, and adapting the configuration files `.travis.yml` and `.codecov.yml` for Travis CI and Codecov services, respectively.

Bringing Jupyter interfaces to command-line based computational mathematics tools makes it possible to interface it with numerous JavaScript libraries, notably for visualization. For example, GAP packages Francy and JupyterViz extend the GAP Jupyter kernel [10] with interactive widgets and plotting tools, which can be tried from their *Binder-ready* repositories.

## Conclusions

In this article, we discuss some of the challenges researchers in computational science and mathematics experience in their everyday work. We focus on making computational exploration and workflows more efficient, more reproducible, and re-usable. We demonstrate the benefits of this approach by showing computational magnetism and computational mathematics use cases. We believe that Project Jupyter and its ecosystem, including JupyterHub and Binder, which allow no-installation browser-based use of notebooks and remote compute resources, can contribute significantly towards more efficient computational workflows, reproducibility and re-usability in science. These conclusions are part of a widespread trend among researchers in the computational community advocating for the use of literate computing – for example using Jupyter – for enhancing reproducible research.

## Acknowledgments

This work was financially supported by the Horizon 2020 European Research Projects OpenDreamKit (676541) and PaNOSC (823852), and the EPSRC Programme grant on Skyrmionics (EP/N032128/1).

## ■ REFERENCES